\documentstyle[12pt,epsfig,here,mine]{article}
\begin{document}

\vspace*{2cm}

\title{THE EXPERIMENTAL LIMITS ON Q-BALL FLUX WITH THE BAIKAL 
       DEEP UNDERWATER ARRAY ``GYRLYANDA''}


\author{
  I.A.BELOLAPTIKOV$^{4,A}$, 
  L.B.BEZRUKOV$^{1,B}$, 
  N.M.BUDNEV$^{2,C}$, 
  E.V.BUGAEV$^{1,D}$,
  ZH.-A.M.DJILKIBAEV$^{1,E}$, 
  G.V.DOMOGATSKY$^{1,F}$, 
  A.A.DOROSHENKO$^{1,G}$, 
  A.M.KLABUKOV$^1$, 
  S.I.KLIMUSHIN$^{1,H}$, 
  L.A.KUZMICHEV$^{3,I}$, 
  A.I.PANFILOV$^1$, 
  YU.V.PARFENOV$^{2,J}$, 
  I.A.SOKALSKI$^{1,K}$\\[8mm] 
}


\address{
{\normalsize
1 - Institute  for  Nuclear  Research,  Russian  Academy  of   Sciences\\}
    117312 60th October Anniversary Prospect 7a, Moscow, Russia\\
{\normalsize
2 - Institute for Applied Physics, Irkutsk State University\\}
    664003 Gagarin boulevard 20, Irkutsk, Russia\\ 
{\normalsize
3 - Institute for Nuclear Physics, Moscow State University\\} 
    119899 Moscow, Russia\\ 
{\normalsize
4 - Joint Institute for Nuclear Research\\}
    141980 Dubna, Moscow region, Russia\\[6mm]
A - belolap@nu.jinr.dubna.su, 
\mbox{B - bezrukov@ms1.inr.ac.ru,} 
\mbox{C - budnev@api.isu.runnet.ru,} 
\mbox{D - bugaev@pcbai10.lpi.msk.su,}
\mbox{E - djilkib@pcbai10.lpi.msk.su,} 
\mbox{F - domogats@pcbai10.lpi.msk.su,} 
\mbox{G - dorosh@pcbai10.lpi.msk.su,}
\mbox{H - klim@pcbai10.lpi.msk.su,} 
\mbox{I - kuz@dec1.npi.msu.su,}
\mbox{J - par@api.isu.runnet.ru,} 
\mbox{K - sokalski@pcbai10.lpi.msk.su}\\[40mm]
{\normalsize
{\bf
Abstract}}
}


\maketitle\abstracts{
Supersymmetric models allow for stable non-topological solitons, Q-balls, 
which can be produced in the early Universe and contribute to dark matter. 
Experimental signature of electrically neutral Q-balls is, in fact, the
same as is expected for superheavy magnetic monopoles catalyzing baryon 
decay. Here we use the upper limits on monopole flux obtained with the deep
underwater Cherenkov array {\sf Gyrlyanda} which operated in the Baikal 
lake in 1984-90 with 267 days of live time to obtain the limit on Q-ball 
flux. The last has been found to be equal to 3.9 $\times$ 10$^{-16}$ 
cm$^{-2}$ sr$^{-1}$ s$^{-1}$ (90$\%$ CL). This result is discussed and 
compared with other restrictions.
}

\pagebreak

\section{Introduction}

Supersymmetric models allow for a new class of objects, which were
named Q-balls~\cite{Coleman85}. These stable non-topological solitons 
can be produced in the early Universe and contribute to dark 
matter~\cite{Coleman85,Kus1,Kus2}. The possible experimental 
signatures of Q-balls were considered 
recently by Kusenko, Kuzmin, 
Shaposhnikov and Tinyakov~\cite{Kuz}. The interactions of Q-balls
with matter were shown in their work to differ essentially on whether 
they are electrically neutral (SENS) or charged (SECS). Q-balls of SENS
type absorb the nuclei with a cross-section

$$
    \sigma \sim 10^{-33}Q^{1/2}(1TeV/m)^{2} cm^{2},\eqno (1)
$$

\noindent
were {\it m} is assumed to be in the range of 0.1 $\div$ 100 TeV and 
{\it Q} is a soliton 
charge (baryon number) which must be not less than $\simeq$
10$^{15}$({\it m}/1TeV)$^{4}$ and may be much greater~\cite{Kus2}. 
The released energy ($\sim$ 1 GeV per nucleon) is emitted in pions which 
(together with their decay products) may become the sources of the Cherenkov 
radiation in a transparent media. 
The Coulomb barrier prevents the absorbtion of 
the incoming nuclei by Q-ball of 
SECS type, which dissipates its energy in collisions with the matter atoms. 
In spite of enormous energy released by SECS passing, e.g., the water ($\sim$ 
100 GeV/cm) it does not result in the Cherenkov light.

So, SENS passing the water media look very much like the monopoles catalyzing 
baryon decay~\cite{ruba} for which the strong experimental flux limits were 
set with the Baikal deep underwater Cherenkov array 
{\sf Gyrlyanda}~\cite{girl}. 
In this short note we give the upper limit on Q-ball (SENS) flux which 
was recalculated from monopole flux limits obtained with {\sf Gyrlyanda} 
in 1984-90. Though no experimental restrictions on Q-ball 
flux have been published by other group so far we compare the 
{\sf Gyrlyanda} results 
with those that can be set using the monopole flux limits. 


\section{Upper limits on monopole flux with the ``Gyrlyanda'' array}

The deep underwater array {\sf Gyrlyanda} was constructed within the framework 
of the Baikal project which is 
devoted to creation of a large scale Cherenkov 
neutrino telescope in Siberian Lake Baikal~\cite{Bai}. 
It operated from April, 1984,
till February, 1990, with live time of 267 days. Being modified 
after each year of operation it consisted of 12--36 PMTs 
placed at the single vertical string at
depthes of 900--1200 m. The distance between upper and lower PMTs was in the 
range of 30--250 m. The number of PMTs involved in the monopole search 
experiment fluctuated from 2 to 24 depending on year with \mbox{$\approx$ 10}
as an averaged value over the whole data taking period.

The detailed description of the {\sf Gyrlyanda} array and monopole search 
experiment
can be found in Ref.~\cite{girl}. The basic idea is as follows: track of 
magnetic monopole catalyzing baryon decays in passage through a water media 
should look as a chain of flashes with a Cherenkov spectrum and, hence, 
objects of such kind can be detected by a short-time excesses of PMTs counting
rate. The main advantage of array which operates in the open water volume is 
that the effective area is determined mainly by catalysis cross-section and 
water 
optical characteristics (in contrast to underground detectors whose effective 
area is limited by their geometrical sizes) and for catalysis cross-section 
$\sigma >$ 10$^{-23}$ cm$^{2}$ may be as large as 10$^{3}$--10$^{5}$ m$^{2}$
even for considerable small single-string array like {\sf Gyrlyanda}.
 
The upper limits set on the monopole flux in Ref.~\cite{girl} lie within 
a range of \mbox{$\simeq$ 10$^{-17}$ --} \mbox{10$^{-14}$ 
cm$^{-2}$ sr$^{-1}$ s$^{-1}$}
depending on monopole velocity $\beta$ and catalysis cross-section 
$\sigma$ which were considered to be within intervals 
of \mbox{10$^{-5}$--10$^{-3}$}
and 10$^{-23}$cm$^{2}$--10$^{-17}$cm$^{2}$, respectively.


\section{Upper limits on Q-ball flux from ``Gyrlyanda'' monopole results}

Q-balls of SECS type passing through the water seem to be not able to generate 
the light flux which would be intensive enough to be detected by an underwater
array. Due to large energy losses it should generate $\sim$ 2 $\times$ 
10$^{4}$ photons per \mbox{1 cm} path via water 
luminiscence~\cite{trof}~\footnote
{
On the one hand it is of $\sim$ 50 times more than the Cherenkov light flux 
generated by relativistic muon which is well detectable object for 
underwater arrays.
But, on the other hand,  due to low velocity of heavy 
Q-ball of SECS type the number of photons generated within some time interval 
is much less
than the corresponding value for the muon. It determines very low value of 
{\sf Gyrlyanda}'s effective area for SECS. 
}. 
It is the same order of magnitude as is expected for SENS type of Q-balls with
the nuclei absorbtion cross-section $\sigma$ $\sim$ 10$^{-24}$ cm$^{2}$ 
and lies under threshold of sensitivity for {\sf Gyrlyanda} array.   

SENS, in contrast to SECS, can produce much more impressive light show moving
in water media. In according to (1), e.g., cross section of nuclei 
absorbtion \mbox{$\sigma >$ 10$^{-23}$ cm$^{2}$} if \mbox{{\it m} = 1 TeV} 
and 
\mbox{{\it Q} $>$ 10$^{20}$}. This means dozens, hundreds, thousands and even
much more (depending on
{\it m} and {\it Q} values) absorbed nuclei per \mbox{1 cm} 
of Q-ball path. Each event 
becomes a source of the Cherenkov radiation 
which is emitted both by pions and their
daughter and grand-daughter particles (muons, {\it e}$^{+}${\it e}$^{-}$ 
pairs etc.).   
The pions multiplicity distribution has not been numerically calculated
so far but should be of the same kind that one for proton-antiproton
annihilation~\cite{kuz2}. Basing on results reported in Ref.~\cite{gol} 
one can expect 2--3 pions on average per each absorbed nucleon. 
Conservative estimates give at least 3 $\times$ 10$^{4}$ Cherenkov photons
with wavelength interval 
300 nm $< \lambda <$ 600 nm 
from each 
\mbox{N -- Q-ball} interaction.   

Thus due to absolutely similar effects which are produced by Q-ball of SENS
type and magnetic monopole catalyzing proton decay in the water media it is
possible to apply the results of Ref.~\cite{girl} to SENS.
If Q-balls of SENS type are responsible for dark matter halo of Galaxy their 
velocities should be $\beta$ $\sim$ 10$^{-3}$~\footnote
{
The sun rotates around the center of Galaxy with a velocity of 
$\beta =$ 7.3 $\times$ 10$^{-4}$.
But there should be some distributions (which are unknown, in fact)
both for Q-balls velocities and their motion directions. This must
cause to some distribution for \mbox{Earth -- Q-balls} relative velocities
which spreads over the range from $\beta \sim$ 0 to $\beta \sim$ 2 
$\times$ 10$^{-3}$.
 
Some fraction of SENS from the halo of Galaxy might loose 
velocity passing, e.g., through giant planet and be gravitationaly
captured by Sun. Such scenario  was considered for GUT monopoles~\cite{fre}
and can be applied to SENS as well. Due to large accumulation time which is 
equal to solar system age ($\sim$ 5 Gyr) it might result 
in remarkable fraction of SENS with velocities of $\beta \sim$ 10$^{-4}$ even 
in spite of negligible part of SENS's kinetic energy which should be lost by 
passing through the matter.  

Nevertheless we assume  here simply $\beta =$ 
10$^{-3}$ for relative \mbox{Earth -- Q-ball} 
velocity because, firstly one is lacking 
in information to use more sophisticated model and, secondly because 
the more detailed analysis is beyond the scope of the present work. 
}.
There are two experimental limits for monopoles catalyzing baryon decay and
moving with velocity $\beta$ = 10$^{-3}$ obtained with {\sf Gyrlyanda} data.
The first limit relates to proton 
decay channel with luminosity L = 1.1 $\times$ 10$^{5}$ Cherenkov photons 
in the spectral interval 300 nm $< \lambda <$ 600 nm, the second one does 
to L = 3.0 $\times$ 10$^{4}$. Both resultes were obtained for 
the effective (averaged for all protons and neutrons forming both
hydrogen and oxygen nuclei) 
catalysis cross-section 
$\sigma$ = 1.9 $\times$ 10$^{-22}$ cm$^{2}$. 
Choosing the
conservative estimation for single \mbox{N -- SENS} interaction luminosity 
L = 3.0 $\times$ 10$^{4}$ we obtain the upper limit for SENS flux: 

$$
       F = 3.9 \times 10^{-16} cm^{-2} sr^{-1} s^{-1} (90\%  CL)\eqno (2)
$$

\noindent
       for nuclei absorbtion cross-section
$$
       \sigma > 1.9 \times 10^{-22} cm^{2}.\eqno (3)
$$

\noindent
For $\beta$ = 10$^{-4}$ and the same values of $\sigma$ the limit is of 
$\simeq$ 10$\%$ more strong.


\section{Discussion}

One should emphasize that for  $\sigma >$ 1.9 $\times$ 10$^{-22}$ cm$^{2}$
the limits are obviously more strong than (2). Moreover for smaller 
cross-sections the restriction
becomes more soft rather smoothly. But there were no numerical calculated
results in Ref.~\cite{girl} for other $\sigma$ values and, therefore one
is not able to obtain the flux limit dependence on cross-section. Following
by the most conservative way let's consider the upper limit (2) to be
constant for all cross-sections which are equal or greater than those
that is determined by (3) and to be not valid for smaller cross-sections.
Using (1) it is easy to obtain the inequality   

$$
    Q^{1/2}(1TeV/m)^{2} > 1.9 \times 10^{11},\eqno (4)
$$

\noindent
which repeats (3) in another therms. The upper limits for SENS flux obtained
here are shown in fig.1 by thick lines 1a, 1b, 1c and 1d. One can see that
for smaller {\it m} the limit is valid for more wide range of {\it Q}. E.g. 
for {\it m} = 0.1 TeV and {\it m} = 100 TeV the flux limit (2) is valid for 
$Q >$ 3.6 $\times$ 10$^{18}$ and $Q >$ 3.6 $\times$ 10$^{30}$, respectively
(lines 1a and 1d). At the same plot the limits wich are caused by dark matter 
density in the galactic halo are shown by four sloping lines (for four
different values of {\it m}). They are resulting from the obvious condition
$\rho_{Q} \le \rho_{DM}$ where $\rho_{DM}$ is galactic halo dark matter 
density 
in the Sun neighboorhood and $\rho_{Q}$ is Q-ball density. Assuming Q-ball
mass to be equal to $M \simeq (4 \pi \sqrt{2}/3)mQ^{3/4}$ (Ref.~\cite{Kuz})  
and $\rho_{DM} \simeq$ 10$^{-2}$ $M_{\odot}$ pc$^{-3}$ 
(see, e.g., Ref.~\cite{dm}), 
we obtain~\footnote 
{
This limit is actual only under assumtion that dark matter is uniformly
distributed along the galactic latitude. Generally saying, this assumption
may be rather far from reality. If so, there may be, e.g., short periods when 
the Earth crosses the galactic areas with 
$\rho_{DM} \gg 10^{-2} M_{\odot} pc^{-3}$ and, on the contrary, long periods
when Earth is inside areas with 
$\rho_{DM} \ll 10^{-2} M_{\odot} pc^{-3}$.
}
$$
     F < 1.5 \times 10^{2} Q^{-3/4} (1TeV/m) cm^{-2} sr^{-1} s^{-1}. \eqno (5)
$$

\noindent
One can see that {\sf Gyrlyanda} flux limits lie below this limit
in rather narrow range of {\it Q} and for small values
of {\it m} only. 

The result which has obtained by the Baksan telescope for 
magnetic monopoles and should be valid for Q-balls of both SENS and SECS
types~\footnote
{
It is our conclusion which seems to be obvious but, strongly saying, it has
to be confirmed by Baksan group.
}
are shown in fig.1 by the thin lines 2a, 2b, 2c. and 2d~\cite{bak}. 
It is only
slightly stronger~\footnote
{
The effective area of the Baksan telescope for Q-balls is much less than
{\sf Gyrlyanda}'s one but Baksan's data taking period is of 20 times longer. 
}
comparing with Baikal result but is valid for much
more wide range of {\it Q}. The Baksan telescope is able to detect SENS
if \mbox{$\sigma \ge$ 5 $\times$ 10$^{-26}$ cm$^{2}$} (compare to (3)).  
The large multi-string deep underwater neutrino telescope {\sf NT-200} is 
currently under construction in the Lake Baikal~\cite{Bai}. The reported
limits which can be set with its data on monopole flux are one or two
orders of magnitude lower than {\sf Gyrlyanda}'s one. These limits will
be able to be applied to SENS flux, too. Due to more phototubes and more 
smart electronics one can expect the less values of $\sigma$ (and, 
consequently, the less values of {\it Q}) for which these more strong limits 
will be valid. 
The experimental upper limits on monopole flux obtained by IMB~\cite{imb},
Kamiokanda~\cite{kam} and MACRO~\cite{MAC} can be also applied to
Q-balls, but resulting limits are still less strong comparing to
both {\sf Gyrlyanda} and Baksan results.
The most strong limit on Q-balls
of SECS type seems to be set by the GUT monopole search
experiment with ``ancient mica''~\cite{mica} and
is equal to \mbox{4 $\times$ 10$^{-19}$ cm$^{-2}$ sr$^{-1}$ s$^{-1}$}. 
If this limit will be confirmed to be valid
for Q-balls of SECS type it will be out of the ability of 
underground/water detectors for a long time.   

Now let's have look at the {\sf Gyrlyanda} limits by another way. 
For {\it m} and {\it Q} above the thick slopping line (fig. 2) the
limit (2) is valid.
The shaded area below the dotted line contains {\it m} and {\it Q} for which
experimental limit (2) is more strong than restriction (5) obtained
from allowed DM density in galactic halo.
So, if one assumes SENS to be entirely responsible for DM density in halo,
one can consider shaded triangle as a region for {\it Q} and {\it m} values
which are excluded by {\sf Gyrlyanda} results. 
One can see that
for presented limit (2) all {\it Q} values have not been excluded for 
{\it m} greater than $\simeq$ 1.5 TeV. To cover the remaining region both 
larger effective
area and ability to detect SENS which absorb the nuclei with smaller
cross-section are neccessary.


\section{Conclusion}

The upper limit on Q-ball flux of SENS type has been set by revising 
the old monopole limits obtained by Baikal deep underwater 
Cherenkov array {\sf Gyrlyanda}:
$F = 3.9 \times 10^{-16} cm^{-2} sr^{-1} s^{-1}$ (90$\%$ CL). It is valid
for nuclei absorbtion cross-section $\sigma > 1.9 \times 10^{-22} cm^{2}$.
It is still above the limit obtained for SENS by Baksan telescope but
can be 
improved with the deep underwater neutrino telescope {\sf NT-200}
which is currently under construction in the Baikal Lake.

The main advantage of underwater arrays operating in the open water volume 
and searching for the objects like magnetic monopoles and Q-balls which
are expected to generate the intensive light flux passing through a water
media is that the effective area is determined mainly by light flux intensity
and water optical characteristics (in contrast to underground detectors whose 
effective area is limited by their geometrical sizes) and may be as large as
10$^{3}$--10$^{5}$ m$^{2}$ even for considerable small arrays.


\section{Acknowledgments}

We have benefited a lot from many stimulating discussions with V.A.Kuzmin.



\pagebreak

\begin{figure}[H]
\centering
\mbox{\epsfig{file=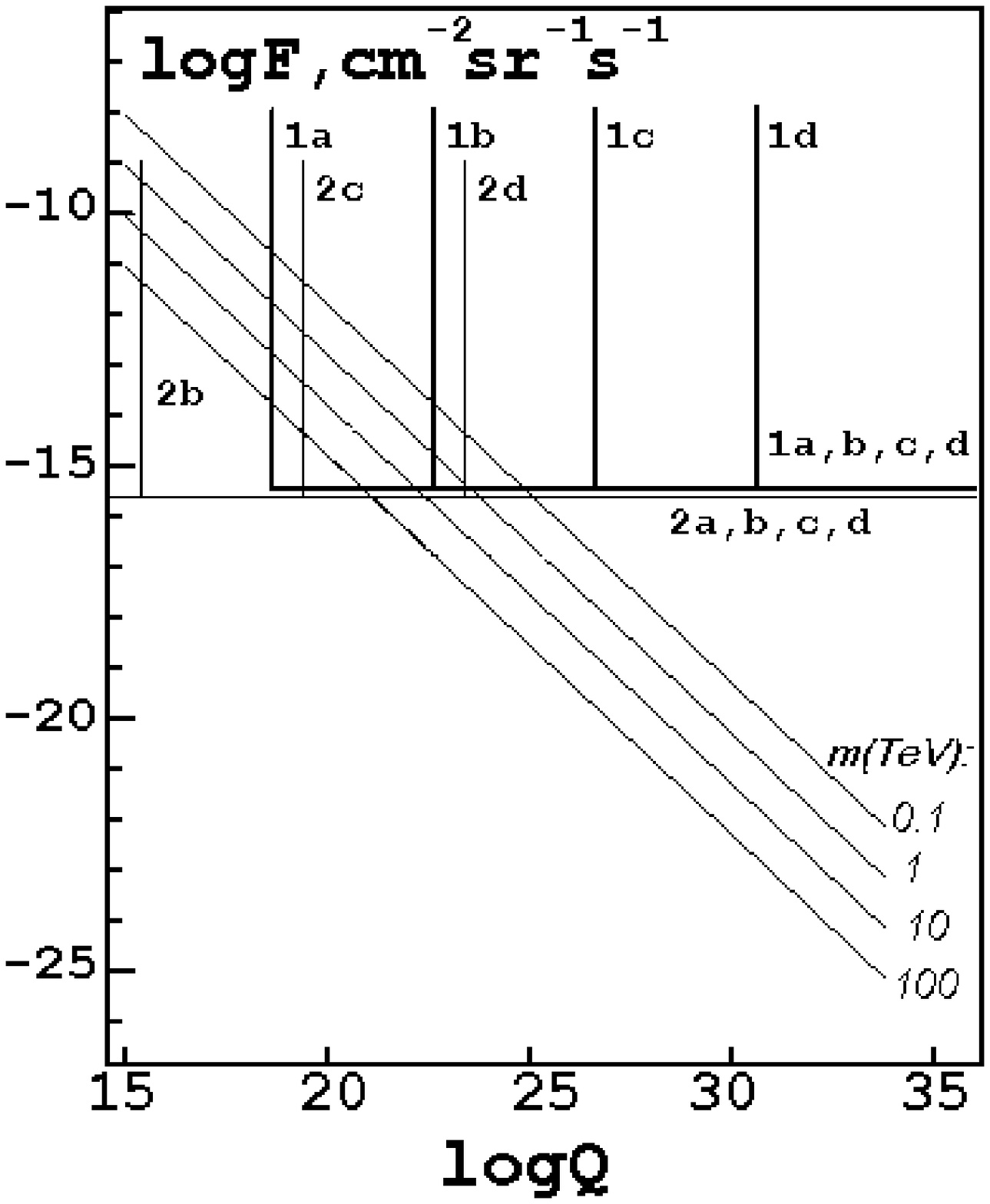}}

\end{figure}

\nopagebreak
\noindent 
{\bf Figure 1:}
Slopping lines: the upper limits on Q-ball flux which are resulting
from condition     
$\rho_{Q} \le \rho_{DM}$ where $\rho_{DM}$ is galactic halo dark matter 
density 
in the Sun neighboorhood and $\rho_{Q}$ is Q-ball density 
($\rho_{DM}$ is assumed to be equal to 10$^{-2}$ $M_{\odot}$ pc$^{-3}$).
Thick lines 1: the upper limits (90$\%$ CL) on Q-ball of SENS type 
flux obtained with Baikal deep underwater
Cherenkov array {\sf Gyrlyanda} for 
{\it m} = 0.1 TeV (1a), 1 TeV (1b),
10 TeV (1c), 100 TeV (1d) (this preprint and Ref.~\cite{girl}).
Thin lines 2: the upper limit for Q-balls (both SENS and SECS type)
obtained with Baksan telescope (90$\%$ CL) for 
{\it m} = 0.1 TeV (2a), 1 TeV (2b),
10 TeV (2c), 100 TeV (2d)~\cite{bak}. See text for the further explanations.
\pagebreak

\begin{figure}[H]
\centering
\mbox{\epsfig{file=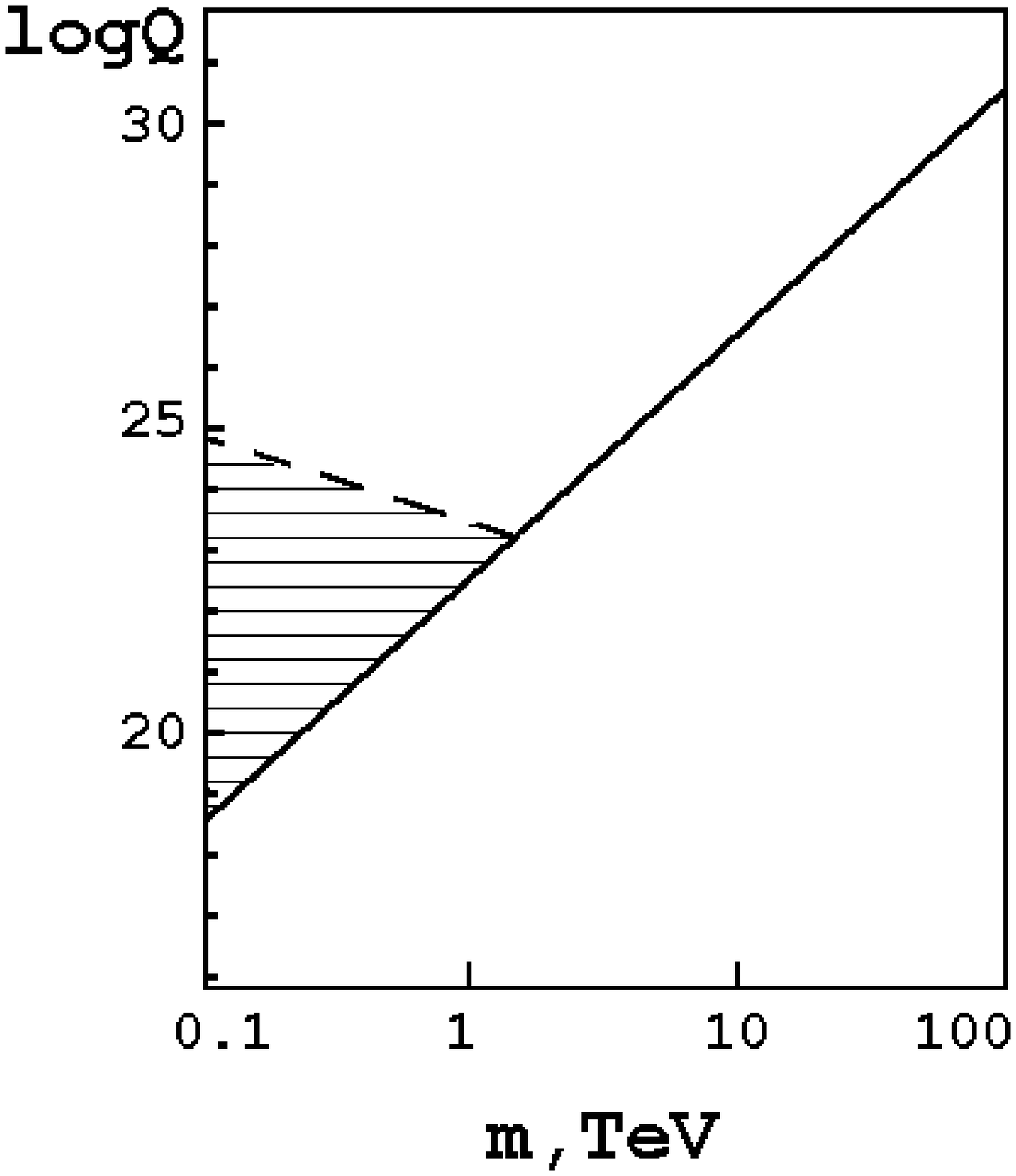}}

\end{figure}

\nopagebreak
\noindent 
{\bf Figure 2:} 
For {\it m} and {\it Q} above the thick slopping line
(including the shaded triangle)
the limit (2) is valid.  
The shaded area below the dotted line contains {\it m} and {\it Q} for which
experimental limit (2) is more strong than restriction (5) obtained
from allowed DM density in galactic halo.
Thus, if one assumes SENS to be entirely responsible for DM density in halo,
one can consider shaded area as a region for {\it Q} and {\it m} values
which are excluded by {\sf Gyrlyanda} results. 

\end{document}